\documentclass[10pt,twocolumn,letterpaper]{article}

\usepackage{cvpr}
\usepackage{times}
\usepackage{epsfig}
\usepackage{graphicx}
\usepackage{amsmath}
\usepackage{amssymb}
\usepackage{multirow}
\usepackage{booktabs}
\usepackage{array}
\usepackage{authblk}
\usepackage{footnote}
\usepackage{bm}
\makesavenoteenv{tabular}

\usepackage[breaklinks=true,bookmarks=false]{hyperref}

\cvprfinalcopy 

\def\cvprPaperID{****} 

\begin{document}

\title{Solution for Authenticity Identification of Typical Target Remote Sensing Images}
\author[1]{Yipeng Lin}
\author[1]{Xinger Li}
\author[1]{{Yang Yang \thanks{Corresponding author: Yang Yang (yyang@njust.edu.cn)}}}
\affil[1]{Nanjing University of Science and Technology}

\maketitle
\begin{abstract}
In this paper, we propose a basic RGB single-mode model based on weakly supervised training under pseudo  labels, which performs high-precision authenticity identification under multi-scene typical target remote sensing images. Due to the imprecision of Mask generation, we divide the task into two sub-tasks: generating pseudo-mask and fine-tuning model based on generated Masks. In generating pseudo masks, we use MM-Fusion as the base model to generate masks for large objects such as planes and ships. By manually calibrating the Mask of a small object such as a car, a highly accurate pseudo-mask is obtained. For the task of fine-tuning models based on generating masks, we use the WSCL model as the base model. It is worth noting that due to the difference between the generated pseudo-Masks and the real Masks, we discard the image feature extractors such as SRM and Noiseprint++ in WSCL, and select the unscaled original image for training alone, which greatly ensures the match between the image and the original label. The final trained model achieved a score of 90.7702 on the test set.

\end{abstract}
\section{Introduction}
Image forgery localization and detection has been a task in the field of forgery for many years. Early work usually focused on specific forgery methods such as splicing  \cite{salloum2018image}, copy-move  \cite{cozzolino2014copy} or removal/inpainting  \cite{li2017localization}. More recently, deep learning-based solutions have been able to identify many different types of operations with greater robustness  \cite{chen2021image,guillaro2023trufor,hu2020span,yang2021s2osc}. By identifying and locating the authenticity of remote sensing images, real remote sensing data can be provided. In the multi-scene-oriented remote sensing image authentication task, it is necessary to recognize the authenticity of typical targets such as aircraft, ships and vehicles on the background of civil airport, sea, land and other images.

We notice that the model is easier to identify the authenticity of large objects such as aircraft and ships, and more difficult to identify small objects such as vehicles. At the same time, it is found that the Mask generated by the small target has a large error, shown in Figure 1. In order to solve the above problems, we designed a method of roughly identifying small targets such as vehicles by manually calibrating masks, and automatically identifying large targets such as aircraft by using MMFusion-IML model \cite{triaridis2024exploring}, and using all the generated pseudo masks, selecting RGB single mode in WSCL model \cite{zhai2023towards} for fine tuning and detection. The MMFusion-IML model uses NoisePrint++ \cite{guillaro2023trufor}, SRM\cite{fridrich2012rich}, Bayar convolution\cite{bayar2016deep} and RGB image modes for Mask positioning, and has accurate positioning ability for larger objects. The original WSCL model uses the weighted Mask generated by SRM, Bayar and RGB, and realizes weakly supervised learning at pixel and picture level through the similarity of Inter-Patch of pictures. Although using RGB image features alone often fails to achieve satisfactory performance in classification tasks with subtle differences in features, integrating multiple modalities to extract latent features from images and ultimately fusing predictions from these modalities tends to outperform individual modalities 
\cite{das2022hate,yang2015auxiliary,yang2018complex,yang2019comprehensive,yang2019deep,yang2021corporate}.

\begin{figure}
\begin{center}
\includegraphics[width=0.45\textwidth]{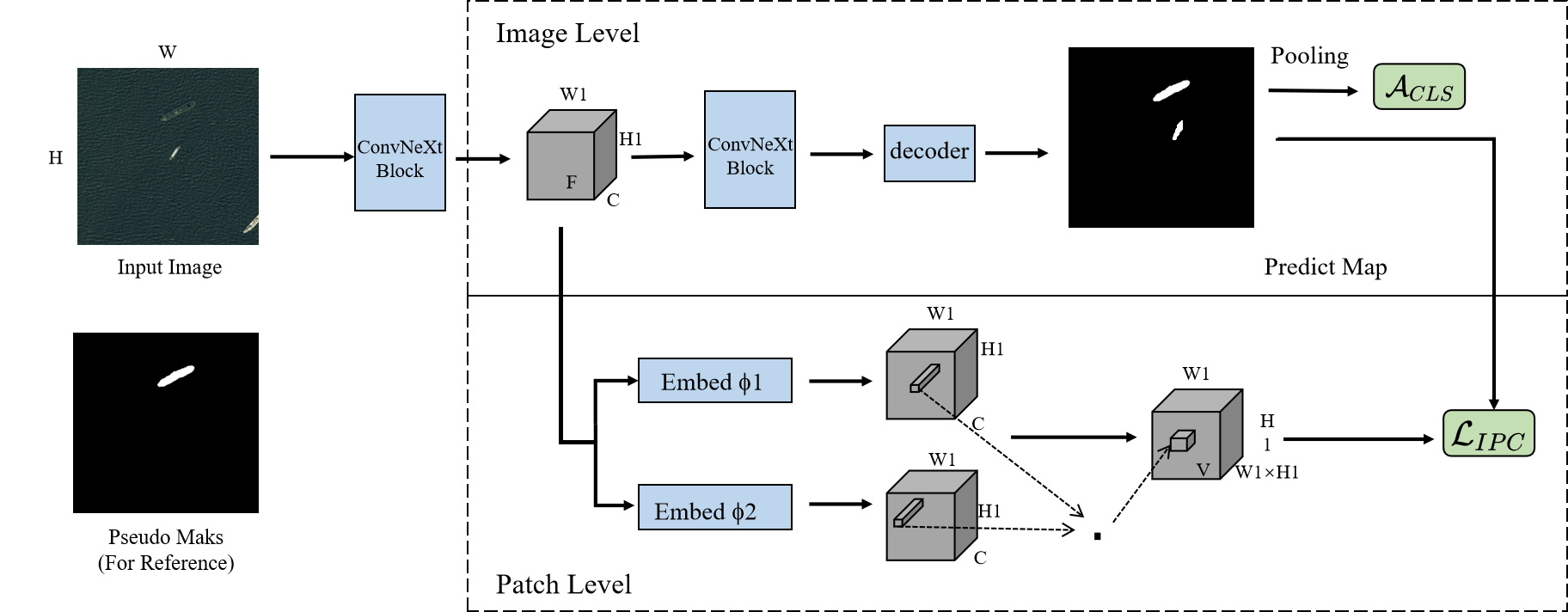}
\end{center}
   \caption{Modified tamper detection model WSCL.}
\label{fig:short}
\end{figure}
Our main contributions are as follows:

$\bullet$ It is found that small targets such as vehicles are difficult to accurately identify, and more accurate pseudo-masks are generated by manual marking.

$\bullet$ Due to the difference between pseudo-mask and GT Mask, WSCL's RGB mode fine-tuning is used separately to avoid the use of multi-mode integration more noise.

$\bullet$ As Resize of the original image will introduce unnecessary noise, only choose to use the complete image to predict.

\section{METHODOLOGY}

The training data set released by this task lacks the mask information corresponding to the tampering position, and through our observation, the size of the tampering information in many pictures is very small, so we need to pay special attention to this part of information in the training process.

However, we do not need to tamper with the precise positioning of the position, we only need to know the approximate position, because our training process does not need the supervision of mask. Therefore, the training process of the model is divided into two parts:

$\bullet$ Generating a more accurate mask of tampering position.

$\bullet$ Using pseudo mask to train picture level and using inter-patch  similarities to train pixel level.
\subsection{GENERATING MASKS}

In order to generate a mask with more accurate location information, we use the Late Fusion version of MMFusion-IML pre-trained model, which achieves the effect of SOTA in image tamper detection and location.

In order to get more data, we randomly cut the data in a batch. When cutting pictures, corresponding masks will be cut, and then the labels of the cut pictures will be reset according to the information of the masks.

In the later experiments, we found that among the pseudo-masks generated by Late Fusion, the Mask location of large-sized objects such as planes and ships was more accurate, but the Mask location of small-sized cars was very poor. In these pictures, background information such as irrelevant trees or roads would be labeled as forged. In order to deal with this problem, we manually locate the box of such pictures, and the Mask before and after calibration is shown in Figure 1.
\begin{figure}
\begin{center}
\includegraphics[width=0.45\textwidth]{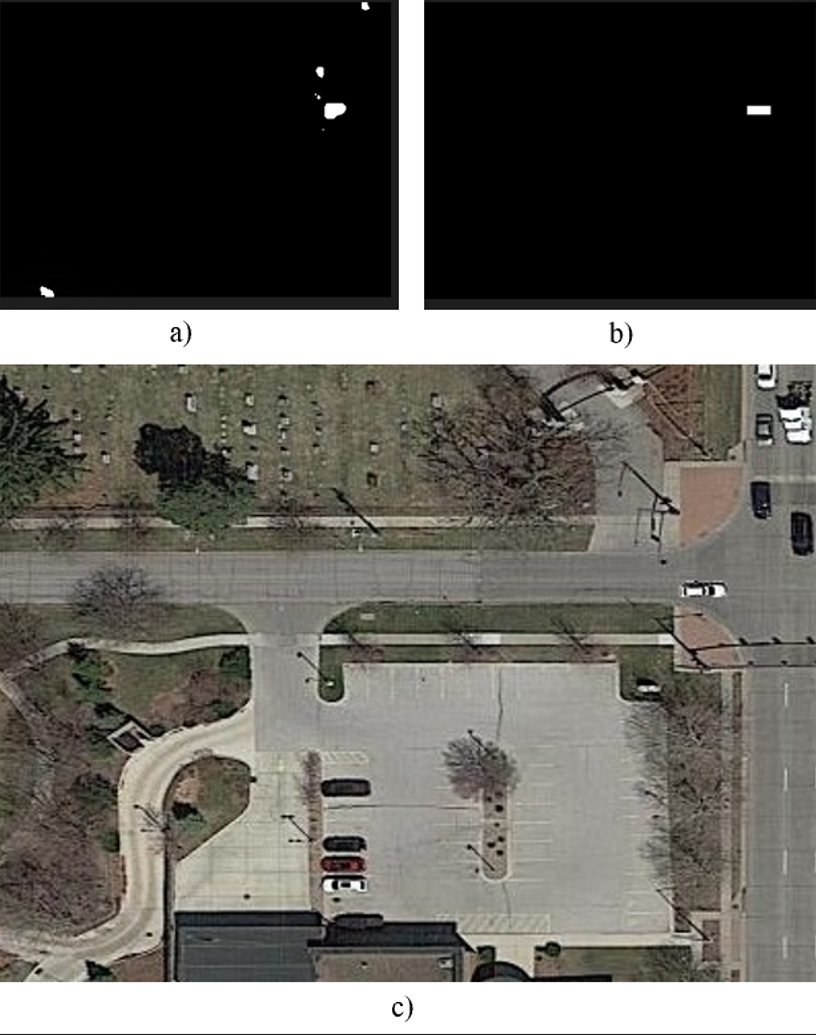}
\end{center}
   \caption{ \textbf{Before and after Mask calibration and the original image.} a) a pseudo-mask generated by MM-Fusion. b) a manually calibrated pseudo-mask. c) the original image that generates the Mask.}
\label{fig:short}
\end{figure}

\subsection{MODEL TRAINING}
We adopted a tamper detection model WSCL [8]  and change the model structure, as shown in Figure 1.

The model uses convnext as the encoder, and the decoder uses a simple 3*3 convolution kernel, The classification header is a mapping of the decoder after the dropout. The training of the model is divided into two parts, one part is Classifier Learning,  another part is Inter-Patch Consistency Learning. The Classifier Learning module focus on features to predict tampered probability. Considering the difference between the generated pseudo-Mask and GT Mask, the potential integration noise of multiple modes, only RGB mode is selected for training in Classifier Learning.

The Inter-Patch Consistency module will extract the intermediate features of the encoder and map them into two images through two different Linear layers. DOT is made between the two images. We compute its dot product similarity against all image patches. 
\begin{equation}
v_{i,j,h,k}=1- \sigma\left (\frac{\phi_{1}(f_{i,j} )*\phi_{2}(f_{h,k} )  }{\sqrt{C} }   \right ) 
\end{equation}

where $ \sigma $ stands for the Sigmoid function. $ \phi_{1} $ and $\phi_{2} $ are two embedding heads realized by MLPs. If patches $f_{i,j} $ and $ f_{h,k} $ share the same forensic characteristic, then $ v_{i,j,h,k} $ = 0, where $ v_{i,j,h,k} $ = 1 indicates  different forensic characteristics. Therefore, authentic images are expected to have all zero consistency volumes, while tampered images should contain at least one location of value 1 in their consistency volumes.

In addition, a comparative learning will be done between the DOT image V and the original image M, where M is the mask location map generated by RGB modes. The Inter-Patch loss function is a simple BCE loss.

\begin{equation}
\mathcal{L}_{IPC} = \frac{1}{H^{'}W^{'}H^{'}W^{'}} \sum_{i,j,h,k} \mathcal{L}_{BCE}(
v_{tgt,i,j,h,k},v_{i,j,h,k})
\end{equation}

After the date of submission, we also make a new version model. In the original code, M and 1-M do DOT operations to generate Vtgt. The author may be aiming at increasing feature generalization. However, after executing the formula with two opposite features, V at the patch level will be updated in the direction that a large number of images will be regarded as fake. We changed it to M and M, namely the default image space is correct most of the time.

\section{EXPERIMENTAL RESULTS}
\textbf{Data.} Data is provided by the competition official. The data is presented in the form of images, ranging in size from 256-2000 pixels. The training data included 1,352 real images and 3,390 fake images. The test data contained 1201 images.

\textbf{Implementation Detail.} The masks of large targets such as aircraft and ships are generated by the pre-trained positioning model MM-Fusion. Model consists of NoisePrint++, Bayer Conv, SRM Conv feature extraction, pre-trained multi-scale image encoder, and pre-trained anomaly decoder. For tamper detection, we use the WSCL model, fine-tuning it on the generated and manually calibrated masks. We used only one NVIDIA RTX 4090 to train the tamper detection model, setting the initial learning rate to 1e-4, batch size to 32, and iteration epoch to 100.

\textbf{Result.} Table 1 shows the performance of our method on the test set. We add our methods step by step in the order of the labels in the table. The experimental results fully confirm the effectiveness of our proposed method. The input of the full original image avoids changes to pixels by resize. The training of RGB single mode significantly improves the accuracy of model detection.

\begin{table}[!htbp]
\caption{Results of our method on the test set.}
\label{table:1}
\begin{center}
\begin{tabular}{ >{\centering\arraybackslash}m{0.05cm} 
>{\centering\arraybackslash}m{6.1cm} 
>{\centering\arraybackslash}m{0.7cm} 
}
 \hline
 \# & Method &  Score \\ 
 \hline
 1 & MM-Fusion detection & $ 70.95+ $\\ 
 2 & WSCL detection & $ 74.34+ $\\ 
 3 & WSCL detection, not Resize image size  & $ 85.29+ $\\ 
 4 & MM-Fusion positioning, WSCL detection, non-resize image size & $ 90.63+$ \\
 5 & MM-Fusion positioning, WSCL detection, non-resize image size, user-defined normalization & $ 90.70+$ \\
 \textbf{6} & \textbf{MM-Fusion positioning, manually calibrated land Mask, WSCL detection, non-resize image size, user-defined normalization, TTA} & $\textbf{90.77+}$ \\ 
\hline
\end{tabular}
\end{center}
\end{table}

We also tries to modify the Backbone of WSCL, namely Resnet, into Convnext-small, Convnext-base and Swin Transformer; Try self-distillation to produce a better pseudo-mask ; Slicing the picture into small patches to make weighted predictions; Slicing the picture and then expand to the original image size to predict, feature fusion of different modalities\cite{yang2023deep,yang2022domfn}, etc., the score is not improved. 

\section{CONCLUSION AND APPLICATIONS}
This report summarizes our solution for a typical target remote sensing image authenticity authentication task for multiple scenarios. Our solution shows that the model has difficulty in locating smaller tampered objects, which can be alleviated to some extent by manual labeling. At the same time, in the case of poor quality of pseudo-mask, training and reasoning using single-mode RGB images are more efficient. Through the above method, the authenticity identification accuracy and AUC area of typical target remote sensing images in multiple scenes can be improved, and the final competition results show the effectiveness of our solution.

The local accuracy of the model with the highest test set score is 95.74\%, and the average processing time of each image is 0.167s.

This paper does not attempt to modify the Decoder of the model and re-extract the features, which involves the modification of the overall structure of the model and a large number of parameters. We only point out the validity of the selected model. In the future, we can pay more attention to the excellent generalization ability of weakly supervised models. Although image tampering is becoming more and more difficult to detect, with only a slightly accurately positioned Mask, we can train an application model that effectively identifies the authenticity of remote sensing images for typical targets in multiple scenes.

{\small
\bibliographystyle{ieee}
\bibliography{egpaper_final}
}

\end{document}